\documentclass[DIV=calc,paper=a4,fontsize=10pt,twocolumn]{scrartcl}

\usepackage{ucs}
\usepackage[utf8x]{inputenc}
\usepackage[english]{babel}

\usepackage{amsmath}
\usepackage{amsfonts}
\usepackage{amssymb}

\usepackage{cite}
\usepackage{multirow}
\usepackage{randtext}

\usepackage{url}
\usepackage[small,labelfont=bf,labelsep=period,justification=raggedright]{caption}
\setcapindent{0pt}

\usepackage{booktabs}
\usepackage[pdftex]{graphicx}
\usepackage[svgnames]{xcolor}
\DeclareGraphicsExtensions{.pdf,.png,.jpg}

\usepackage{hyperref}
\hypersetup{
	pdfauthor={P. Widera and N. Krasnogor},
	pdftitle={Protein Models Comparator: Scalable Bioinformatics Computing on the
	Google App Engine Platform},
	pdfkeywords={protein structure comparison, protein model comparison,
	bioinformatics, web application, web service, distributed computing,
	cloud computing, google app engine}
}

\usepackage[protrusion=true,expansion=true]{microtype}
\usepackage{fix-cm}

\usepackage{sectsty}
\allsectionsfont{
	\usefont{OT1}{phv}{b}{n} %
}

\usepackage{titlesec}
\titlespacing*{\section}{0pt}{*4}{*1}
\titlespacing*{\subsection}{0pt}{*3}{*1}

\usepackage{lastpage}
\usepackage{fancyhdr}
\usepackage{titling}

\newcommand{\hr}{
	\color{DarkGoldenrod}
	\rule{\linewidth}{1pt}
}

\pretitle{
	\vspace{-1em}
	\begin{flushleft}
		\hr
		\vskip1em
		\color{DarkRed}
		\fontsize{34}{34}
		\usefont{OT1}{phv}{b}{n}
		Protein Models Comparator \\
		\vskip0.5em
}
\title{}
\posttitle{
		\fontsize{26}{26}
		\usefont{OT1}{phv}{b}{n}
 		Scalable Bioinformatics Computing on the Google App Engine Platform
		\par
	\end{flushleft}
	\vskip 0.5em
}

\preauthor{
 	\begin{flushleft}
	 	\large
		\lineskip 0.5em
		\usefont{OT1}{phv}{b}{sl}
		\color{DarkGreen}
}
\author{
	Paweł Widera and Natalio Krasnogor
}
\postauthor{
		\vskip0.75em
		\footnotesize
		\usefont{OT1}{phv}{m}{sl}
		\color{Black}
		School of Computer Science, University of Nottingham, UK \\
		\textbf{Contact:} \randomize{natalio.krasnogor@nottingham.ac.uk}
	\end{flushleft}
	\hr
}

\date{}

\usepackage{lettrine}
\newcommand{\initial}[1]{%
 	\lettrine[lines=3,lhang=0.3,nindent=0em]{
		\color{DarkGoldenrod}{\textsf{#1}}}{}}

\begin{document}

\twocolumn[

\maketitle
\vspace{-5.5em}

\begin{@twocolumnfalse}

\paragraph{Background:}

The comparison of computer generated protein structural models is an important
element of protein structure prediction. It has many uses including model
quality evaluation, selection of the final models from a large set of candidates
or optimisation of parameters of energy functions used in template-free
modelling and refinement. Although many protein comparison methods are available
online on numerous web servers, they are not well suited for large scale model
comparison: (1) they operate with methods designed to compare actual proteins,
not the models of the same protein, (2) majority of them offer only a single
pairwise structural comparison and are unable to scale up to a required order of
thousands of comparisons. To bridge the gap between the protein and model
structure comparison we have developed the Protein Models Comparator (pm-cmp).
To be able to deliver the scalability on demand and handle large comparison
experiments the pm-cmp was implemented ``in the cloud''.

\paragraph{Results:}

Protein Models Comparator is a scalable web application for a fast distributed
comparison of protein models with RMSD, GDT\_TS, TM-score and Q-score measures.
It runs on the Google App Engine cloud platform and is a showcase of how the
emerging PaaS (Platform as a Service) technology could be used to simplify the
development of scalable bioinformatics services. The functionality of pm-cmp is
accessible through API which allows a full automation of the experiment
submission and results retrieval. Protein Models Comparator is a free software
released under the Affero GNU Public Licence and is available with its source
code at: \url{http://www.infobiotics.org/pm-cmp}

\paragraph{Conclusions:}

This article presents a new web application addressing the need for a
large-scale model-specific protein structure comparison and provides an insight
into the GAE (Google App Engine) platform and its usefulness in scientific
computing.

\hr \vskip2.5em
\end{@twocolumnfalse}
]

\thispagestyle{fancy}


\initial{P}rotein structure comparison seems to be most successfully applied to
the functional classification of newly discovered proteins. As the evolutionary
continuity between the structure and the function of proteins is strong, it is
possible to infer the function of a new protein based on its structural
similarity to known protein structures. This is, however, not the only
application of structural comparison. There are several aspects of protein
structure prediction (PSP) where robust structural comparison is very important.

The most common application is the evaluation of models. To measure the quality
of a model, the predicted structure is compared against the target native
structure. This type of evaluation is performed on a large scale during the CASP
experiment (Critical Assessment of protein Structure Prediction), when all
models submitted by different prediction groups are ranked by the similarity to
the target structure. Depending on the target category, which could be either a
template-based modelling (TBM) target or a free modelling (FM) target, the
comparison emphasis is put either on local similarity and identification of well
predicted regions or global distance between the model and the native structure
\cite{Zhang2008, Cozzetto2008a, Kryshtafovych2009}.

The CASP evaluation is done only for the final models submitted by each group.
These models have to be selected from a large set of computer generated
candidate structures of unknown quality. The most promising models are commonly
chosen with the use of clustering techniques. First, all models are compared
against each other and then, split into several groups of close similarity
(clusters). The most representative elements of each cluster (e.g. cluster
centroids) are selected as final models for submission \cite{Shortle1998,
Zhang2004c}.

The generation of models in the free modelling category, as well as the process
of model refinement in both FM and TBM categories, requires a well designed
protein energy function. As it is believed that the native structure is in a
state of thermodynamic equilibrium and low free energy, the energy function is
used to guide the structural search towards more native-like structures.
Ideally, the energy function should have low values for models within small
structural distance to the native structure, and high values for the most
distinct and non-protein-like models. To ensure such properties, the parameters
of energy functions are carefully optimised on a training set of models for
which the real distances to the native structures are precomputed
\cite{Zhang2003, Rohl2004, Widera2010, Zhang2010}.

\subsection{Model comparison vs. protein alignment}

All these three aspects of prediction: evaluation of models quality, selection
of the best models from a set of candidates and the optimisation of energy
functions, require a significant number of structural comparisons to be made.
However, this comparisons are not made between two proteins, but between
two protein models that are structural variants of the same protein and are composed
of the same set of atoms. Because of that, the alignment between the atoms is
known a priori and is fixed, in contrast to comparison between two different
proteins where the alignment of atoms usually has to be found before scoring the
structural similarity.

\begin{figure}[h]
\centering
\includegraphics[width=\columnwidth]{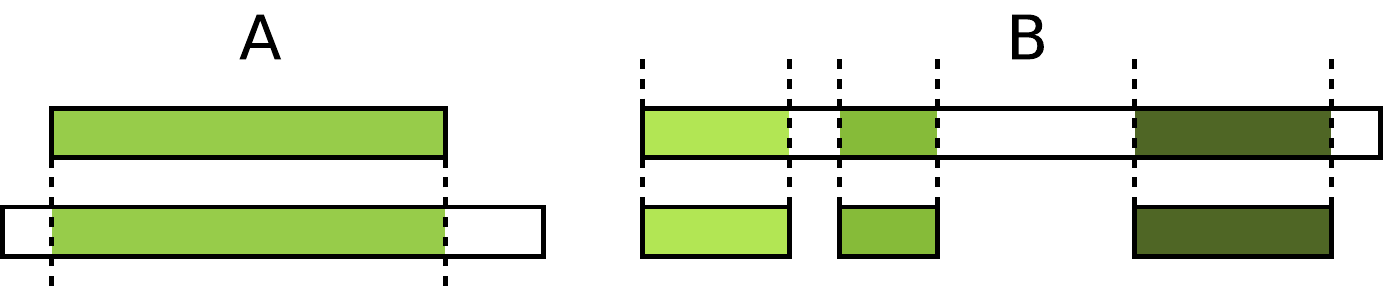}
\caption{
	\textbf{Matching common residues between two structures.} There are two common cases when
	number of residues differs between the structures: (A) some residues at the
	beginning/end of a protein sequence were not located in the crystallography
	experiment and (B) structure was derived from templates that did not cover the
	entire protein sequence. In both cases pm-cmp performs a comparison
	using the maximum common subset of residues.
}
\label{fig:matching}
\end{figure}

Even though searching for optimal alignment is not necessary in model
comparison, assessing their similarity is still not straightforward. Additional
complexity is caused in practice by the incompleteness of models. For example,
many CASP submitted models contain the atomic coordinates for just a subset of
the protein sequence. Often even the native structures have several residues
missing as the X-ray crystallography experiments not always locate all of them.
As the model comparison measures operate only on the structures of equal length,
a common set of residues have to be determined for each pair of models before
the comparison is performed (see Figure~\ref{fig:matching}). It should be noted
that this is not an alignment in the traditional sense but just a matching
procedure that selects the residues present in both structures.

\begin{figure*}[t]
\centering
\includegraphics[width=0.9\textwidth]{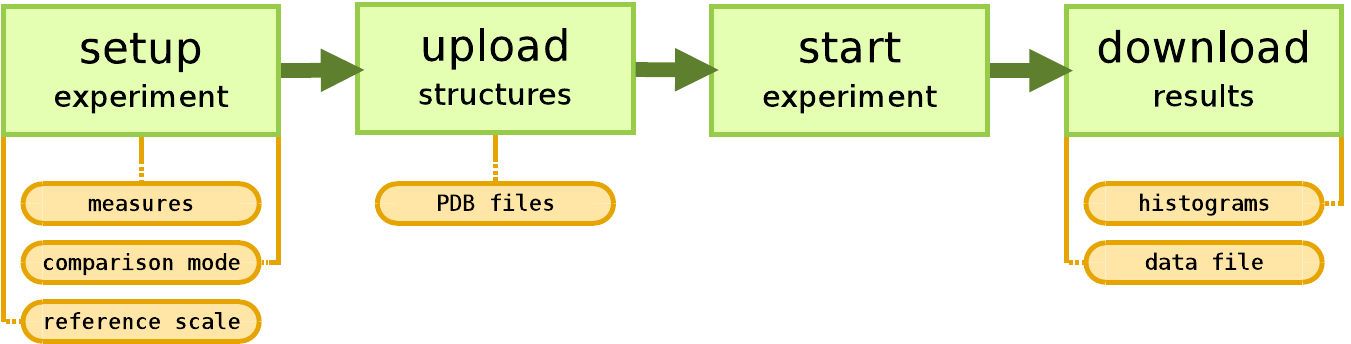}
\caption{
	\textbf{Application control flow.} The interaction with a user is divided into
	4 steps: setup of the experiment options, upload of the structural models,
	start of the computations and finally download of the results when ready.
}
\label{fig:control_flow}
\end{figure*}

\subsection{Comparison servers}

Although many protein structure comparison web services are already available
online, they are not well suited for models comparison. Firstly, they do not
operate on a scale needed for such a task. Commonly these methods offer a simple
comparison between two structures (1:1) or in the best case, a comparison
between a single structure and a set of known structures extracted from the
Protein Data Bank (1:PDB). While what is really needed is the ability to compare
a large number of structures either against a known native structure (1:N) or
against each other (N:N). Secondly, the comparison itself is done using just a
single comparison method, which may not be reliable enough for all the cases
(types of proteins, sizes etc.).

An exception to this is the ProCKSI server \cite{Barthel2007} that uses several
different comparison methods and provides 1:N and N:N comparison modes. However,
it operates with methods designed to compare real proteins, not the models
generated in the process of PSP, and therefore it lacks the ability to use a
fixed alignment while scoring the structural similarity. Also the high
computational cost of these methods makes large-scale comparison experiments
difficult without a support of grid computing facilities (see our previous work
on this topic \cite{Folino2009, Shah2010}).

The only server able to perform a large-scale model-specific structural
comparison we are aware of, is the infrastructure implemented to support the
CASP experiment \cite{Kryshtafovych2009a}. This service, however, is only
available to a small group of CASP assessors for the purpose of evaluation of
the predictions submitted for a current edition of CASP. It is a closed and
proprietary system that is not publicly available neither as an online server
nor in a form of a source code. Due to that, it cannot be freely used,
replicated or adapted to the specific needs of the users. We have created the
Protein Models Comparator (pm-cmp) to address these issues.

\subsection{Google App Engine}

We implemented pm-cmp using the Google App Engine (GAE) \cite{URL_GAE}, a
recently introduced web application platform designed for scalability. GAE
operates as a cloud computing environment providing Platform as a Service
(PaaS), and removes the need to consider physical resources as they are
automatically scaled up as and when required. Any individual or a small team
with enough programming skills can build a distributed and scalable application
on GAE without the need to spend any resources on the setup and maintenance of
the hardware infrastructure. This way, scientist freed from tedious
configuration and administration tasks can focus on what they do best, the
science itself.

GAE offers two runtime environments based on Python or Java. Both environments
offer almost identical set of platform services, they only differ in maturity as
Java environment has been introduced 12 months after first preview of the Python
one. The environments are well documented and frequently updated with new
features. A limited amount of GAE resources is provided for free and is enough to
run a small application. This limits are consequently decreased with each release
of the platform SDK (Software Development Kit) as the stability and performance
issues are ironed out. There are no set-up costs and all payments are based on
the daily amount of resources (storage, bandwidth, CPU time) used above the free
levels.

In the next sections we describe the overall architecture and functionality of
our web application, exemplify several use cases, present the results of the
performance tests, discuss the main limitations of our work and point out a few
directions for the future.


\section{Implementation}

The pm-cmp application enables users to set up a comparison experiment with a
chosen set of similarity measures, upload the protein structures and download the
results when all comparisons are completed. The interaction between pm-cmp and
the user is limited to four steps presented in Figure~\ref{fig:control_flow}.

\subsection{Application architecture}

The user interface (UI) and most of the application logic was implemented in
Python using the web2py framework \cite{URL_WEB2PY}. Because web2py provides an
abstraction layer for data access, this code is portable and could run outside
of the GAE infrastructure with minimal changes. Thanks to the syntax brevity of
the Python language and the simplicity of web2py constructs the pm-cmp
application is also very easy to extend. For visualisation of the results the UI
module uses Flot \cite{URL_FLOT}, a JavaScript plotting library.

The comparison engine was implemented in Groovy using Gaelyk \cite{URL_GAELYK},
a small lightweight web framework designed for GAE. It runs in Java Virtual
Machine (JVM) environment and interfaces with the BioShell java library
\cite{Gront2008} that implements a number of structure comparison methods. We
decided to use Groovy for the ease of development and Python-like programming
experience, especially that a dedicated GAE framework (Gaelyk) already existed.
We did not use any of the enterprise level Java frameworks such as Spring,
Stripes, Tapestry or Wicket as they are more complex (often require an
sophisticated XML-based configuration) and were not fully compatible with GAE,
due to specific restriction of its JVM. However, recently a number of
workarounds have been introduced to make some of this frameworks usable on GAE.

\begin{figure}[hbt]
\centering
\includegraphics[width=\columnwidth]{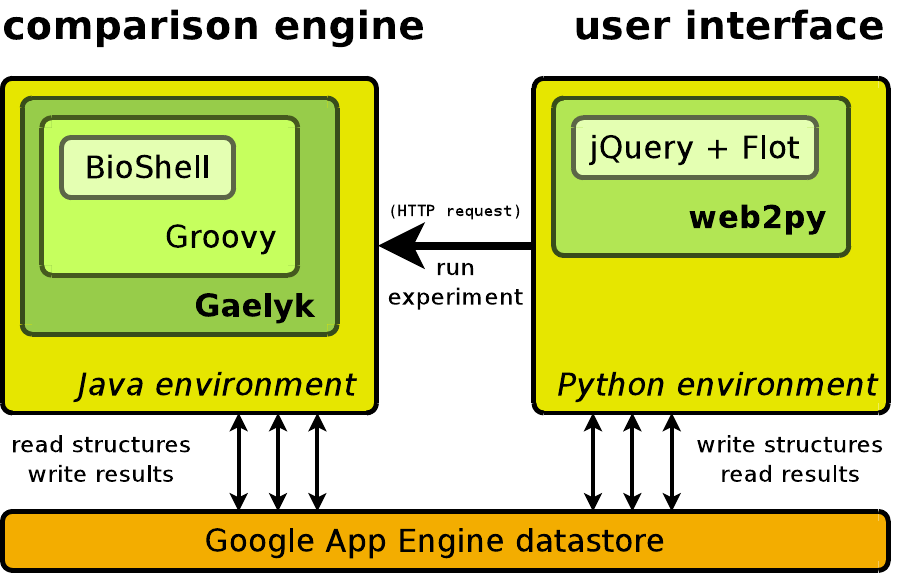}
\caption{
	\textbf{Protein Models Comparator architecture.} The application GUI was
	implemented in the GAE Python environment. It guides the user through
	the setup of an experiment and then sends HTTP request to the comparison
	engine to start the computations. The comparison engine was implemented in
	the GAE Java environment.
}
\label{fig:architecture}
\end{figure}

The communication between the UI module and the comparison engine is done with
the use of HTTP request. The request is sent when all the structures have been
uploaded and the experiment is ready to start (see Figure~\ref{fig:architecture}).
The comparison module organises all the computational work required for the
experiment into small tasks. Each task, represented as HTTP request, is put into
a queue and later automatically dispatched by GAE according to the defined
scheduling criteria.

\subsection{Distribution of tasks}

\begin{figure}[ht]
\centering
\includegraphics[width=\columnwidth]{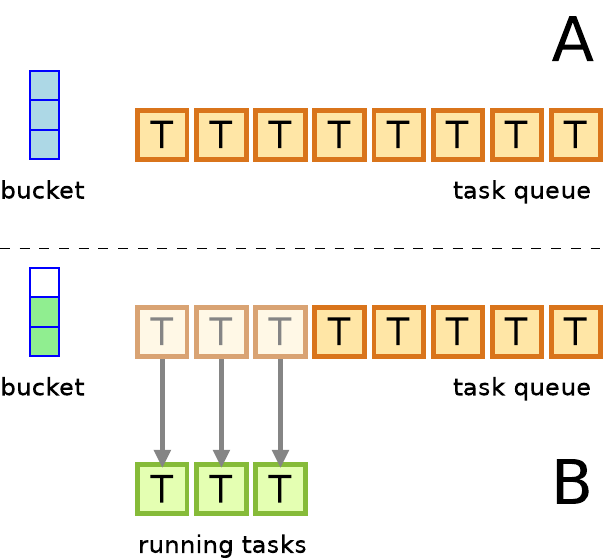}
\caption{
	\textbf{Task queue management on Google App Engine.} A) 8 tasks has been added to a
	queue. The token bucket is full and has 3 tokens. B) Tokens are used to run 3
	tasks and the bucket is refilled with 2 new tokens.
}
\label{fig:queue}
\end{figure}

The task execution on GAE is scheduled with a token bucket algorithm that has
two parameters: a bucket size and a bucket refill rate. The number of tokens in
the bucket limits the number of tasks that could be executed at a given time
(see Figure~\ref{fig:queue}). The tasks that were executed in parallel run on the
separate instances of the application in the cloud. New instances are automatically
created when needed and deleted after a period of inactivity which enables the
application to scale dynamically to the current demand.

Our application uses tasks primarily to distribute the computations, but also for
other background activities like deletion of uploaded structures or old
experiments data. The computations are distributed as separate structure vs.
structure comparison tasks. Each task reads the structures previously written
to the datastore by the UI module, performs the comparison and stores back the
results. This procedure is slightly optimised with a use of the GAE memcache
service and each time a structure is read for the second time it is served
from a fast local cache instead of being fetched from the slower distributed
datastore. Also to minimise the number of datastore reads all selected measures
are computed together in a single task.

The comparison of two structures starts with a search for the common $C_\alpha$
atoms. Because the comparison methods require both structures to be equal in
length, a common atomic denominator is used in the comparison. If required, the
total length of the models is used as a reference for the similarity scores,
so that the score of a partial match is proportionally lower than the score of a
full length match. This approach makes the comparison very robust, even for
models of different size (as long as they share a number of atoms).

\section{Results}

\begin{figure}[b!]
\centering
\includegraphics[width=\columnwidth]{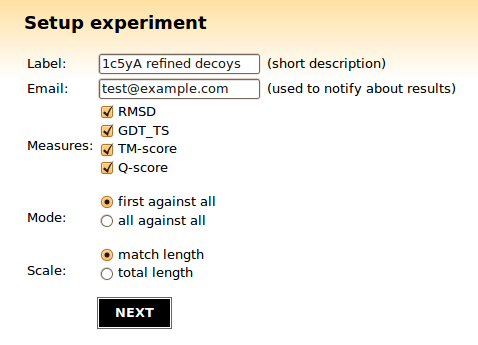}
\caption{
	\textbf{Experiment setup screen.} To set up an experiment the user has to
	choose a label for it, optionally provide an e-mail address (if she wants to be
	notified about the experiment status), select one or more comparison measures,
	and choose the comparison mode (1:N or N:N) and the reference scale. }
\label{fig:setup}
\end{figure}

The pm-cmp application provides a clean interface to define a comparison
experiment and upload the protein structures. In each experiment the user can
choose which measures and what comparison mode (1:N or N:N) should be used (see
Figure~\ref{fig:setup}). Currently, four structure comparison measures are
implemented: RMSD, GDT\_TS \cite{Zemla2003}, TM-score \cite{Zhang2004b} and
Q-score \cite{Hardin2002}. These are the main measures used in evaluation of
CASP models.

Additionally, a user can choose the scale of reference for GDT\_TS and TM-score.
It could be the number of matching residues or the total size of the structures
being compared. It changes the results only if the models are incomplete. The
first option is useful when a user is interested in the similarity score
regardless of the number of residues used in comparison. For example, she
submits incomplete models containing only coordinates of residues predicted with
high confidence and wants to know how good these fragments are alone. On the
other hand, a user might want to take into account all residues in the
structures being compared, not just the matching ones. For that, she would use
the second option where the similarity score is scaled by the length of the
target structure (in 1:N comparison mode) or by the length of the shorter
structure from a pair being compared (in N:N comparison mode). This way a short
fragment with a perfect match will have a lower score than a less perfect
full-length match.

After setting up the experiment, the next step is the upload of models. This is
done with the use of Flash to allow multiple file uploads. The user can track
the progress of the upload process of each file and the number of files left in
the upload queue. When the upload is finished a user can start the computations,
or if needed, upload more models.

\begin{figure}[htb]
\centering
\includegraphics[width=\columnwidth]{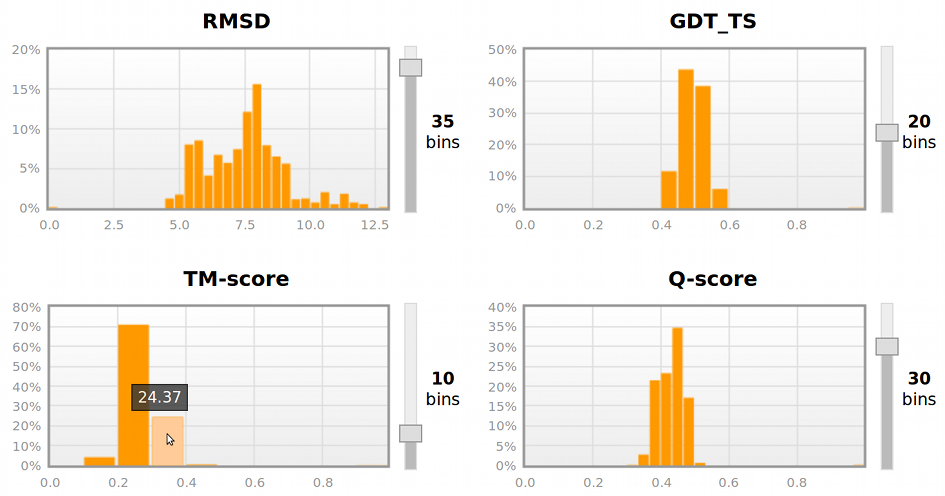}
\caption{
	\textbf{Example of distribution plots.} For a quick visual assessment of models diversity
	the results of comparison are additionally presented as histograms of the
	similarity/distance values.
}
\label{fig:distribution}
\end{figure}

The current status of recently submitted experiments is shown on a separate
page. Instead of checking the status there, a user can provide an e-mail address
on experiment setup to be notified when the experiment is finished. The results
of the experiment are presented in a form of interactive histograms showing for
each measure the distribution of scores across the models (see
Figure~\ref{fig:distribution}). Also a raw data file is provided for download
and possible further analysis (e.g. clustering). In case of errors the user is
notified by e-mail and a detailed list of problems is given. In most cases
errors are caused by inconsistencies in the set of models, e.g. lack of common
residues, use of different chains, mixing models of different proteins or
non-conformance to the PDB format. Despite the errors, the partial results are
still available and contain all successfully completed comparisons.

\begin{table*}[bt]
\resizebox{\textwidth}{!} {
\begin{tabular}{lcp{6.5cm}p{3cm}}
	\toprule
	\textbf{URL} & \textbf{Method} & \textbf{Parameters} & \textbf{Return} \\
	\midrule
	\multirow{4}{*}{\texttt{/experiments/setup}} & \multirow{4}{*}{POST} &
	\texttt{label} - string & \multirow{4}{*}{303 Redirect} \\
	& & \texttt{measures} - subset of [\textit{RSMD}, \textit{GDT\_TS}, \textit{TM-score}, \textit{Q-score}] & \\
	& & \texttt{mode} - \textit{first against all} or \textit{all against all} & \\
	& & \texttt{scale} - \textit{match length} or \textit{total length} & \\
	\texttt{/experiments/structures/[id]} & POST & \texttt{file} - multipart/form-data encoded file & HTML link to the uploaded file \\
	\texttt{/experiments/start/[id]} & GET & - & 200 OK \\
	\texttt{/experiments/status/[id]} & GET & - & status in plain text \\
	\texttt{/experiments/download/[id]} & GET & - & results file \\
	\bottomrule
\end{tabular}
}
\caption{Description of the RESTful interface of pm-cmp.}
\label{tab:api}
\end{table*}

There are three main advantages of pm-cmp over the existing online services for
protein structure comparison. First of all, it can work with multiple structures
and run experiments that may require thousands of pairwise comparisons.
Secondly, these comparisons are performed correctly, even if some residues are
missing in the structures, thanks to the residue matching mechanism. Thirdly, it
integrates several comparison measures in a single service giving the users an
option to choose the aspect of similarity they want to test their models with.

\subsection{Application Programming Interface (API)}

As Protein Models Comparator is build in the REST (REpresentational State
Transfer) architecture, it is easy to access programmatically. It uses standard
HTTP methods (e.g. GET, POST, DELETE) to provide services and communicates back
the HTTP response codes (e.g. 200 - OK, 404 - Not Found) along with the content.
By using the RESTful API summarised in Table~\ref{tab:api}, it is possible to
set up an experiment, upload the models, start the computations, check the
experiment status and download the results file automatically. We provide
\texttt{pm-cmp-bot.py}, an example of a Python script that uses this API to
automate the experiment submission and results retrieval. As we wanted to keep
the script simple and readable, the handling of connection problems is limited
to the most I/O intensive upload part and in general the script does not retry
on error, verify the response, etc. Despite of that, it is a fully functional
tool and it was used in several tests described in the next section.

\subsection{Performance tests}

To examine the performance of the proposed architecture we ran a 48h test in
which a group of beta testers ran multiple experiments in parallel at different
times of a day. As a benchmark we used the models generated by I-TASSER \cite{Wu2007},
one of the top prediction methods in the last three editions of CASP.

From each set containing every 10th structure from the I-TASSER simulation
timeline we selected the top $n$ models, i.e. the closest to the native by
means of RMSD. The number of models was chosen in relation to the protein length
to obtain one small, two medium and one large size experiment as shown in
Table~\ref{tab:benchmark_data}. The smallest experiment was four times
smaller the the large one and two times smaller than the medium one.

\begin{table}[htb]
\resizebox{\columnwidth}{!} {
\begin{tabular}{ccccc}
	\toprule
	\textbf{protein} & 1b72A & 1kviA & 1egxA & 1fo5A \\
	\midrule
	\textbf{(models*length)} & (350x49) & (500x68) & (300x115) & (800x85) \\
	\textbf{total size} & 17150 & 34000 & 34500 & 68000 \\
	\bottomrule
\end{tabular}
}
\caption{Four sets of protein models used in the performance benchmark
(available for download on the pm-cmp website).}
\label{tab:benchmark_data}
\end{table}

We observed a very consistent behaviour of the application, with a relative
absolute median deviation of the total experiment processing time smaller than
10\%. The values reported in Table~\ref{tab:benchmark_performance} show the
statistics for 15 runs per each of the four sets of models. The task queue
rate was set to 4/s with a bucket size of 10. Whenever execution of two experiments
overlapped, we accounted for this overlap by subtracting the waiting time from
the execution time, so that the time spent in a queue while the other experiment
was still running was not counted. Using GAE 1.2.7 we were able to run about 30
experiments per day staying within the free CPU quota.

\begin{table}[!h]
\resizebox{\columnwidth}{!}{
\begin{tabular}{rrrrrrr}
	\toprule
	& & & \multicolumn{4}{c}{\textbf{processing time[s]}} \\
	\cmidrule(r){4-7}
	\textbf{protein} & \textbf{models} & \textbf{length} &
	\textbf{median} & \textbf{mad$^*$} & \textbf{min} & \textbf{max} \\
	\midrule
	1b72A & 350 & 49 & 178 & 17 & 108 & 272 \\
	1egxA & 300 & 115 & 195 & 17 & 125 & 274 \\
	1kviaA & 500 & 68 & 236 & 16 & 203 & 406 \\
	1fo5A & 800 & 85 & 369 & 33 & 307 & 459 \\
	\bottomrule
\end{tabular}
}
\caption*{
	\scriptsize *) \textbf{mad} (median absolute deviation) =
	$median_i(|x_i-median(X)|)$
}
\caption{Results of the performance benchmark.}
\label{tab:benchmark_performance}
\end{table}

To test the scalability of pm-cmp we ran additional two large experiments with
approximately 2500 comparisons each (using GAE 1.3.8). We used the models
generated by I-TASSER again: 2500 models for [PDB:1b4bA] (every 5th structure
from the simulation timeline) and 70 models for [PDB:2rebA2] (top models from
every 10th structure sample set). The results of 11 runs per set are summarised
in Table~\ref{tab:2500cmp}. All runs were separated by a 15 minutes inactivity
time, to allow GAE to bring down all active instances. Thus each run activated
the application instances from scratch, instead of reusing instances activated
by the previous run. Because the experiments did not overlap and due to the use
of more mature version of the GAE platform, the relative absolute median
deviation was much lower than in the first performance benchmark and did not
exceed 3.5\%.

\begin{table}[htb]
\resizebox{\columnwidth}{!} {
\begin{tabular}{rrrrrrrr}
	\toprule
	& & & \multicolumn{4}{c}{\textbf{processing time[s]}} \\
	\cmidrule(r){4-7}
	\textbf{experiment} & \textbf{models} & \textbf{length} &
	\textbf{median} & \textbf{mad} & \textbf{min} &	\textbf{max} \\
	\midrule
	1b4bA (1:N 2501 cmp) & 2500 & 71 & 838.00 & 25.00 & 746 & 903 \\
	2reb\_2 (N:N 2415 cmp) & 70 & 60 & 854.00 & 29.00 & 731 & 958 \\
	\bottomrule
\end{tabular}
}
\caption{Performance for large number of comparisons.}
\label{tab:2500cmp}
\end{table}

To relate the performance of our application to the performance of the
comparison engine executed locally we conducted another test. This time we
followed a typical CASP scenario and we evaluated 308 server submitted models
for the CASP9 target T0618 ([PDB:3nrhA]). The comparison against the target
structure was performed with the use of the \texttt{pm-cmp-bot} and two times
were measured: experiment execution time (as in previous test) and the total
time used by \texttt{pm-cmp-bot} (including upload/download times). The
statistics of 11 runs are reported in Table~\ref{tab:performance_vs_local}. As
the experiments were performed in 1:N mode the file upload process took a
substantial 30\% of the total time. The local execution of the comparison engine
on a machine with Intel P8400 2.26GHz (2 core CPU) was almost 5 times
slower than the execution in the cloud. We consider this to be a significant
speed up, especially having in mind the conservative setting of the task queue
rate (4/s while GAE allows a maximum of 100/s). Our preliminary experiments with
GAE 1.4.3 showed that the speedup possible with the queue rate of 100 tasks per
second is at least an order of magnitude larger.

\begin{table}[htb]
\resizebox{\columnwidth}{!} {
\begin{tabular}{ccrrrr}
	\toprule
	& & \multicolumn{4}{c}{\textbf{processing time[s]}} \\
	\cmidrule(r){3-6}
	\textbf{platform} & \textbf{time} &
	\textbf{median} & \textbf{mad} & \textbf{min} & \textbf{max} \\
	\midrule
	GAE & total & 135 & 4 & 127 & 146 \\
	GAE & execution & 89 & 2 & 86 & 97 \\
	local & execution &  413 & 8 & 394 & 422 \\
	\bottomrule
\end{tabular}
}
\caption{Performance compared to local execution.}
\label{tab:performance_vs_local}
\end{table}


\section{Discussion}

The pm-cmp application is a convenient tool performing a comparison of a
set of protein models against a target structure (e.g. in model quality assessment
or optimisation of energy functions for PSP) or against each other (e.g. in
selection of the most frequently occurring structures). It is also an
interesting showcase of a scalable scientific computing on the Google App Engine
platform. To provide more inside on the usefulness of GAE in bioinformatics
applications in general, we discuss below the main limitations of our approach,
possible workarounds and future work.

\subsection{Response time limit}

A critical issue in implementing an application working on GAE was to keep the
response time to each HTTP request below the 30s limit. This is why the division
of work into small tasks and extensive use of queues was required. However, this
might be no longer critical in the recent releases of GAE 1.4.x which allow the
background tasks to run 20 times longer. In our application, where a single
pairwise comparison with all methods never took longer than 10s, the task
execution time was never an issue. The bottleneck was the task distribution
routine. As it was not possible to read more than 1000 entities from a datastore
within the 30s time limit, our application was not able to scale up above the
1000 comparisons per experiment. However, GAE 1.3.1 introduced the mechanism of
cursors to tackle this very problem. That is, when a datastore query is
performed its progress can be stored in a cursor and resumed later. Our code
distribution routine simply call itself (by adding self to the task queue) just
before the time limit is reached and continue the processing in the next cycle.
This way our application could scale up to thousands of models. However, as it
currently operates within the free CPU quota limit, we do not allow very large
experiments online yet. For practical reasons we set the limit to 5000
comparisons. This allows us to divide the daily CPU limit between several
smaller experiments, instead of having it all consumed by a single large
experiment. In the future we would like to monitor the CPU usage and adjust the
size of the experiment with respect to the amount of the resources left each
day.

\subsection{Native code execution}

Both environments available on GAE are build on interpreted languages. This is
not an issue in case of a standard web applications, however in scientific
computing the efficiency of the code execution is very important (especially in
the context of response time limits mentioned above). A common practice of
binding these languages with fast native modules written in C/C++ is
unfortunately not an option on GAE. No arbitrary native code can be run on the
GAE platform, only the pure Python/Java modules. Although Google extends the
number of native modules available on GAE it is rather unlikely that we will see
anytime soon modules for fast numeric computation such as NumPy. For that reason
we implemented the comparison engine on the Java Virtual Machine, instead of
using Python.

\subsection{Bridging Python and Java}

Initially we wanted to run our application as a single module written in Jython
(implementation of Python in Java) that runs inside a Java Servlet and then
bridge it with web2py framework to combine Python's ease of use with the
numerical speed of the JVM. However, we found that this is not possible without
mapping all GAE Python API calls made by web2py framework to its Java API
correspondents. As the amount of work needed to do that exceeded the time we had
to work on the project we attempted to join these two worlds differently. We
decided to implement it as two separate applications, each in its own
environment, but sharing the same datastore. This was not possible as each GAE
application can access only its own datastore. We had to resort to the mechanism
of versions. It was designed to test a new version of an application while the
old one is still running. Each version is independent from the others in terms
of the used environment and they all share the same datastore. This might be
considered to be a hack and not a very elegant solution but it worked exactly as
intended; we end up with two separate modules accessing the same data.

\subsection{Handling large files}

There is a hard 1MB limit on the size of a datastore entity. The dedicated
Blobstore service introduced in GAE 1.3.0 makes the upload of large files
possible but as it was considered experimental and did not provide at first an
API to read the blob content, we decided not to use it. As a consequence we
could not use a simple approach of uploading all experiment data in a single
compressed file. Instead, we decided to upload the files one by one directly to
the datastore, since a single protein structure file is usually much smaller
than 1MB. To make the upload process easy and capable of handling hundreds of
files, we used the Uploadify library \cite{URL_UPLOADIFY} which combines
JavaScript and Flash to provide a multi-files upload and progress tracking.
Although since GAE 1.3.5 it is now possible to read the content of a blob, the
multiple file decompression still remains a complex issue because GAE lacks a
direct access to the file system. It would be interesting to investigate in the
future if a task cycling technique (used in our distribution routine) could be
used to tackle this problem.

\subsection{Vendor lock-in}

Although the GAE code remains proprietary, the software development kit (SDK)
required to run the GAE stack locally is a free software licensed under the Apache
Licence 2.0. Information contained in the SDK code allowed the creation of two
alternative free software implementations of the GAE platform:
AppScale \cite{URL_APPSCALE} and TyphoonAE \cite{URL_TYPHOONAE}.
The risk of vendor lock-in is therefore minimised as the same application code
could be run outside of the Google's platform if needed.

\subsection{Comparison to other cloud platforms}

GAE provides an infrastructure that automates much of the difficulties related to
creating scalable web applications and is best-suited for small and medium-sized
applications. Applications that need high performance computing, access to
relational database or direct access to operating system primitives might be
better suited for more generic cloud computing frameworks.

There are two major competitors to the Google platform. Microsoft's Azure
Services are based on the .NET framework and provide a less constrained
environment, but require to write more low level code and do not guarantee
scalability. Amazon Web Services are a collection of low-level tools that provide
Infrastructure as a Service (IaaS), that is storage and hardware. Users can
assign a web application to as many computing units (instances of virtual
machines) as needed. They also receive complete control over the machines, at the
cost of requiring maintenance and administration. Similarly to Microsoft's cloud,
it does not provide automated scalability, so it is clearly a trade-off between
access at a lower and unconstrained level and the scalability that has to be
implemented by the user. Additionally, both these platforms are fully paid
services and do not offer free/start-up resources.

\section{Conclusions}

Protein Models Comparator is filling the gap between commonly offered online
simple 1:1 protein comparison and the non-public proprietary CASP large-scale
evaluation infrastructure. It has been implemented using Google App Engine
platform that offers automatic scalability on the data storage and the task
execution level.

In addition to a friendly user web interface, our service is accessible through
REST-like API that allows full automation of the experiments (we provide an
example script for remote access). Protein Models Comparator is a free software,
which means anyone can study and learn from its source code as well as extend it
with his own modifications or even set up clones of the application either on
GAE or using one of the alternative platforms such as AppScale or TyphoonAE.

Although GAE is a great platform for prototyping as it eliminates the need to
set up and maintain the hardware, provides the resources on demand and
automatic scalability, the task execution limit makes it suitable only for
highly parallel computations (i.e. the ones that could be split into small
independent chunks of work). Also a lack of direct disk access and inability to
execute the native code restricts the possible uses of GAE. However, looking
back at the history of changes it seems likely that in the future GAE platform
will become less and less restricted. For example, the long running background
tasks had been on the top of the GAE project roadmap \cite{URL_GAE_ROADMAP} and
recently the task execution limit was raised in GAE 1.4.x making the platform
more suitable for scientific computations.

\section{Acknowledgements}

We would like to thank the fellow researchers who kindly devoted their time to
testing the pm-cmp: E. Glaab, J. Blakes, J. Smaldon, J. Chaplin, M. Franco, J.
Bacardit, A.A. Shah, J. Twycross and C. García-Martínez.

This work was supported by the Engineering and Physical Sciences Research
Council [EP/D061571/1]; and the Biotechnology and Biological Sciences Research
Council [BB/F01855X/1].

\bibliographystyle{ieeetr-url}
\footnotesize
\onecolumn
\bibliography{references}

\begin{thebibliography}{10}

\bibitem{Zhang2008}
Y.~Zhang, ``{P}rogress and challenges in protein structure prediction,'' {\em
  Current Opinion in Structural Biology}, vol.~18, pp.~342--348, Jun 2008.
\newblock \href {http://dx.doi.org/10.1016/j.sbi.2008.02.004}
  {\path{doi:10.1016/j.sbi.2008.02.004}}.

\bibitem{Cozzetto2008a}
D.~Cozzetto, A.~Giorgetti, D.~Raimondo, and A.~Tramontano, ``{T}he {E}valuation
  of {P}rotein {S}tructure {P}rediction {R}esults,'' {\em Molecular
  Biotechnology}, vol.~39, no.~1, pp.~1--8, 2008.
\newblock \href {http://dx.doi.org/10.1007/s12033-007-9023-6}
  {\path{doi:10.1007/s12033-007-9023-6}}.

\bibitem{Kryshtafovych2009}
A.~Kryshtafovych and K.~Fidelis, ``{P}rotein structure prediction and model
  quality assessment,'' {\em Drug Discovery Today}, vol.~14, pp.~386--393, Apr.
  2009.
\newblock \href {http://dx.doi.org/10.1016/j.drudis.2008.11.010}
  {\path{doi:10.1016/j.drudis.2008.11.010}}.

\bibitem{Shortle1998}
D.~Shortle, K.~T. Simons, and D.~Baker,
  \href{http://www.pnas.org/content/95/19/11158.full} {``{C}lustering of
  low-energy conformations near the native structures of small proteins,''}
  {\em Proceedings of the National Academy of Sciences of the United States of
  America}, vol.~95, pp.~11158--11162, Sept. 1998 [cited 2010-09-21].

\bibitem{Zhang2004c}
Y.~Zhang and J.~Skolnick, ``{SPICKER}: {A} clustering approach to identify
  near-native protein folds,'' {\em J. Comput. Chem.}, vol.~25, no.~6,
  pp.~865--871, 2004.
\newblock \href {http://dx.doi.org/10.1002/jcc.20011}
  {\path{doi:10.1002/jcc.20011}}.

\bibitem{Zhang2003}
Y.~Zhang, A.~Kolinski, and J.~Skolnick, ``{TOUCHSTONE} {II}: {A} {N}ew
  {A}pproach to {A}b {I}nitio {P}rotein {S}tructure {P}rediction,'' {\em
  Biophys. J.}, vol.~85, pp.~1145--1164, Aug. 2003.
\newblock \href {http://dx.doi.org/10.1016/S0006-3495(03)74551-2}
  {\path{doi:10.1016/S0006-3495(03)74551-2}}.

\bibitem{Rohl2004}
C.~A. Rohl, C.~E.~M. Strauss, K.~M.~S. Misura, and D.~Baker, ``{P}rotein
  {S}tructure {P}rediction {U}sing {R}osetta,'' in {\em Numerical Computer
  Methods, Part D} (L.~Brand and M.~L. Johnson, eds.), vol.~Volume 383 of {\em
  Methods in Enzymology}, pp.~66--93, Academic Press, Jan. 2004.
\newblock \href {http://dx.doi.org/10.1016/S0076-6879(04)83004-0}
  {\path{doi:10.1016/S0076-6879(04)83004-0}}.

\bibitem{Widera2010}
P.~Widera, J.~Garibaldi, and N.~Krasnogor, ``{GP} challenge: evolving energy
  function for protein structure prediction,'' {\em Genetic Programming and
  Evolvable Machines}, vol.~11, pp.~61--88, March 2010.
\newblock \href {http://dx.doi.org/10.1007/s10710-009-9087-0}
  {\path{doi:10.1007/s10710-009-9087-0}}.

\bibitem{Zhang2010}
J.~Zhang and Y.~Zhang, ``{A} {N}ovel {S}ide-{C}hain {O}rientation {D}ependent
  {P}otential {D}erived from {R}andom-{W}alk {R}eference {S}tate for {P}rotein
  {F}old {S}election and {S}tructure {P}rediction,'' {\em PLoS ONE}, vol.~5,
  p.~e15386, Oct. 2010.
\newblock \href {http://dx.doi.org/10.1371/journal.pone.0015386}
  {\path{doi:10.1371/journal.pone.0015386}}.

\bibitem{Barthel2007}
D.~Barthel, J.~D. Hirst, J.~Blazewicz, and N.~Krasnogor, ``{P}ro{CKSI}: {A}
  {D}ecision {S}upport {S}ystem for {P}rotein ({S}tructure) {C}omparison,
  {K}nowledge, {S}imilarity and {I}nformation,'' {\em BMC Bioinformatics},
  vol.~8, no.~1, p.~416, 2007.
\newblock \href {http://dx.doi.org/10.1186/1471-2105-8-416}
  {\path{doi:10.1186/1471-2105-8-416}}.

\bibitem{Folino2009}
G.~Folino, A.~Shah, and N.~Kransnogor, ``{O}n the storage, management and
  analysis of (multi) similarity for large scale protein structure datasets in
  the grid,'' in {\em 22nd IEEE International Symposium on Computer-Based
  Medical Systems (CBMS 2009)}, pp.~1--8, aug 2009.
\newblock \href {http://dx.doi.org/10.1109/CBMS.2009.5255328}
  {\path{doi:10.1109/CBMS.2009.5255328}}.

\bibitem{Shah2010}
A.~Shah, G.~Folino, and N.~Krasnogor, ``{T}oward {H}igh-{T}hroughput,
  {M}ulticriteria {P}rotein-{S}tructure {C}omparison and {A}nalysis,'' {\em
  IEEE Transactions on NanoBioscience}, vol.~9, pp.~144--155, jun 2010.
\newblock \href {http://dx.doi.org/10.1109/TNB.2010.2043851}
  {\path{doi:10.1109/TNB.2010.2043851}}.

\bibitem{Kryshtafovych2009a}
A.~Kryshtafovych, O.~Krysko, P.~Daniluk, Z.~Dmytriv, and K.~Fidelis,
  ``{P}rotein structure prediction center in {CASP}8,'' {\em Proteins},
  vol.~77, no.~S9, pp.~5--9, 2009.
\newblock \href {http://dx.doi.org/10.1002/prot.22517}
  {\path{doi:10.1002/prot.22517}}.

\bibitem{URL_GAE}
\href{http://code.google.com/appengine/} {``{G}oogle {A}pp {E}ngine,''}
  [online, cited 2009-11-06].

\bibitem{URL_WEB2PY}
M.~D. Pierro.
\newblock \href{http://www.web2py.com/} {``web2py web framework,''} [online,
  cited 2009-11-06].

\bibitem{URL_FLOT}
O.~Laursen.
\newblock \href{http://code.google.com/p/flot/} {``{F}lot - {J}avascript
  plotting library for j{Q}uery,''} [online, cited 2011-02-18].

\bibitem{URL_GAELYK}
M.~Overdijk and G.~Laforge.
\newblock \href{http://gaelyk.appspot.com/} {``{G}aelyk - lightweight {G}roovy
  toolkit for {G}oogle {A}pp {E}ngine,''} [online, cited 2009-11-06].

\bibitem{Gront2008}
D.~Gront and A.~Kolinski, ``{U}tility library for structural bioinformatics,''
  {\em Bioinformatics}, vol.~24, pp.~584--585, Feb. 2008.
\newblock \href {http://dx.doi.org/10.1093/bioinformatics/btm627}
  {\path{doi:10.1093/bioinformatics/btm627}}.

\bibitem{Zemla2003}
A.~Zemla, ``{LGA}: a method for finding 3{D} similarities in protein
  structures,'' {\em Nucl. Acids Res.}, vol.~31, no.~13, pp.~3370--3374, 2003.
\newblock \href {http://dx.doi.org/10.1093/nar/gkg571}
  {\path{doi:10.1093/nar/gkg571}}.

\bibitem{Zhang2004b}
Y.~Zhang and J.~Skolnick, ``{S}coring function for automated assessment of
  protein structure template quality,'' {\em Proteins: Structure, Function, and
  Bioinformatics}, vol.~57, pp.~702--710, Jan. 2004.
\newblock \href {http://dx.doi.org/10.1002/prot.20264}
  {\path{doi:10.1002/prot.20264}}.

\bibitem{Hardin2002}
C.~Hardin, M.~P. Eastwood, M.~Prentiss, Z.~Luthey-Schulten, and P.~G. Wolynes,
  ``{F}olding funnels: {T}he key to robust protein structure prediction,'' {\em
  Journal of Computational Chemistry}, vol.~23, no.~1, pp.~138--146, 2002.
\newblock \href {http://dx.doi.org/10.1002/jcc.1162}
  {\path{doi:10.1002/jcc.1162}}.

\bibitem{Wu2007}
S.~Wu, J.~Skolnick, and Y.~Zhang, ``{A}b initio modeling of small proteins by
  iterative {TASSER} simulations.,'' {\em BMC Biol}, vol.~5, p.~17, May 2007.
\newblock \href {http://dx.doi.org/10.1186/1741-7007-5-17}
  {\path{doi:10.1186/1741-7007-5-17}}.

\bibitem{URL_UPLOADIFY}
R.~Garcia and T.~Nickels.
\newblock \href{http://www.uploadify.com/} {``{U}ploadify - a multiple file
  upload plugin for j{Q}uery,''} [online, cited 2009-11-06].

\bibitem{URL_APPSCALE}
C.~Krintz, C.~Bunch, N.~Chohan, J.~Chohan, N.~Garg, M.~Hubert, J.~Kupferman,
  P.~Lakhina, Y.~Li, G.~Mehta, N.~Mostafa, Y.~Nomura, and S.~H. Park.
\newblock \href{http://appscale.cs.ucsb.edu/} {``{A}pp{S}cale - open source
  implementation of the {G}oogle {A}pp {E}ngine,''} [online, cited 2010-09-21].

\bibitem{URL_TYPHOONAE}
T.~Rodaebel and F.~Glanzner.
\newblock \href{http://code.google.com/p/typhoonae/} {``{T}yphoon{AE} -
  environment to run {G}oogle {A}pp {E}ngine ({P}ython) applications,''}
  [online, cited 2010-09-21].

\bibitem{URL_GAE_ROADMAP}
\href{http://code.google.com/appengine/docs/roadmap.html} {``{G}oogle {A}pp
  {E}ngine project roadmap,''} [online, cited 2010-09-21].

\end{thebibliography}

\end{document}